\documentclass[aps,pra,twocolumn,longbibliography ]{revtex4-1}
\usepackage{graphicx}
\usepackage{dcolumn}
\usepackage{bm}
\usepackage{color}
\usepackage{amsmath}
\usepackage{amssymb}
\usepackage{setspace}

\usepackage{adjustbox}  

\begin{document}
\title{Thermal effects of  the quantum states generated from the isomorphs of PPKTP crystal}
\author{Rui-Bo Jin$^{1,2}$}
\email{jrbqyj@gmail.com}
\author{Guo-Qun Chen$^{1}$}
\author{Fabian Laudenbach$^{3,4}$}
\author{Shengmei Zhao$^{2}$}
\author{Pei-Xiang Lu$^{1}$}

\affiliation{$^{1}$Laboratory of Optical Information Technology, Wuhan Institute of Technology, Wuhan 430205, China}
\affiliation{$^{2}$Key Lab of Broadband Wireless Communication and Sensor Network Technology, Nanjing University of Posts and Telecommunications, Ministry of Education}
\affiliation{$^{3}$Security $\&$  Communication Technologies, Center for Digital Safety $\&$  Security, \\ AIT Austrian Institute of Technology GmbH, Donau-City-Str. 1, 1220 Vienna, Austria}
\affiliation{$^{4}$ Quantum Optics, Quantum Nanophysics and Quantum Information, Faculty of Physics,  University of Vienna, Boltzmanngasse 5, 1090 Vienna, Austria}

\begin{abstract}
We theoretically and numerically investigate the temperature-dependent properties of the biphotons generated from four isomorphs of periodically poled $\mathrm{KTiOPO_4}$ (PPKTP): i.e., PPRTP,  PPKTA, PPRTA and PPCTA.
It is discovered that the first type of group-velocity-matched (GVM) wavelength is decreased by 6.4, 1.2, 8.9, 25.6 and 6.3 nm, while the phase-matched wavelength is decreased by 4.4, -0.4, -1.2, 29.1 and 59.5 nm  for PPKTP, PPRTP, PPKTA, PPRTA and PPCTA, respectively, when the temperature is increased from 20$\,^{\circ}\mathrm{C}$ to 120$\,^{\circ}\mathrm{C}$.
Although the maximal spectral purity of the heralded single photons is not changed at different temperature, the Hong-Ou-Mandel (HOM) interference shows different patterns due to a shift of the joint spectral amplitude.
These thermal effects are very important for precise control of the quantum state for the future applications in quantum information processing, for example, in quantum interference or spectroscopy.
\end{abstract}

\maketitle

\section{\emph{Introduction}}
Quantum state engineering using nonlinear optical crystals in a spontaneous parametric down-conversion (SPDC) process is a very important technique for the study of quantum information processing (QIP)\cite{Grice2001,  Edamatsu2011}. Periodically poled $\mathrm{KTiOPO_4}$  (PPKTP) is one of the widely used crystals, because it satisfies the group-velocity-matched (GVM) condition and therefore can be used to prepare spectrally pure state at telecom wavelength \cite{Evans2010, Gerrits2011, Eckstein2011, Jin2013OE, Zhou2013, Zhou2013PRA, Bruno2014OE, Li2015, Jin2016SR, Laudenbach2016, Chen2017, Jin2018Optica}. Recently, it was discovered that spectrally pure states can also be generated from  four isomorphs of the PPKTP, i.e., periodically poled RTP ($\mathrm{RbTiOPO_4}$), KTA ($\mathrm{KTiOAsO_4}$), RTA ($\mathrm{RbTiOAsO_4}$) and CTA ($\mathrm{CsTiOAsO_4}$), with a  general form of  \emph{M}TiO\emph{X}O$_4$ with \{\emph{M} = K, Rb, Cs\} and \{\emph{X} = P, As (for \emph{M}=Cs only)\} \cite{Jin2016PRAppl, Laudenbach2017, Cheng1994}. It was found that these isomorphs still retain the desirable properties of their parent PPKTP, namely  high spectral purity (over 0.8) with wide frequency tunability (more than 400 nm) at a variety of wavelengths (from 1300 nm to 2100 nm). These isomorphs  may provide more and better choices for quantum state engineering at telecom wavelengths.

In the previous study in Refs. \cite{Jin2016PRAppl, Laudenbach2017}, the properties of the quantum state generated from the \emph{M}TiO\emph{X}O$_4$ are investigated only at room temperature, i.e., 20$\,^{\circ}\mathrm{C}$. However, their temperature-dependent properties are still unexplored.
Temperature is an important parameter for these quasi-phase matched (QPM) crystals, because the phase-matching condition in the QPM crystals is mainly controlled by temperature in experiment. Temperature is also the key to precisely control the quantum state in many QIP applications, e.g., in fine tuning of the  quantum wave function \cite{Tischler2015} and entanglement \cite{Fedrizzi2009NJP}, in spectroscopy \cite{Whittaker2017} and quantum interference \cite{Branczyk2011}.
For future applications of these isomorphs, it is therefore very important and necessary to investigate the temperature-dependent properties of quantum states generated by these materials.
With this motivation, in this work we study the thermal effect on spectrally-pure-state generated from the periodically poled \emph{M}TiO\emph{X}O$_4$  crystals.

This paper is organized as follows: Section I is the introduction part. In section II, we study the GVM wavelength of the five crystals as a function of temperature.  In section III, the phase matched wavelength as a function of temperature will be explored. In section IV, we study the thermal effect on the joint spectral amplitude (JSA) of the biphotons generated from the isomorphs. Based on different JSAs, different Hong-Ou-Mandel (HOM) interference patterns are also investigated. Then, we provide some discussions in section V and summarize the paper in section VI.

\section{\emph{Thermal effect on the GVM wavelength }}
\begin{table*}[tbp]
  \centering
  \begin{adjustbox}{max width=\textwidth}
  \begin{tabular}{c|ccccc}
  \hline \hline
   Name &PPKTP               &PPRTP                 &PPKTA                 &PPRTA                  &  PPCTA    \\
   Composition &$\mathrm{KTiOPO_4}$ &$\mathrm{RbTiOPO_4}$  &$\mathrm{KTiOAsO_4}$  &$\mathrm{RbTiOAsO_4}$  &$\mathrm{CsTiOAsO_4}$ \\
   \hline
    $\lambda_{GVM1}$  at 20$\,^{\circ}\mathrm{C}$ (nm)   &1584.6   &1643.2         &1680.9        &1786.6                 &1972.5 \\

    $\Delta \lambda_{GVM1}$   20-120$\,^{\circ}\mathrm{C}$  (nm)  &6.4   &1.2       &8.9          &25.6         &6.3 \\
  \hline
  $\lambda_{GVM2}$  at 20$\,^{\circ}\mathrm{C}$  (nm)          &1225.2   &1282.0     &1288.1       &1379.7         &1577.2\\
  $\Delta \lambda_{GVM2}$   20-120$\,^{\circ}\mathrm{C}$ (nm)   &7.3     &-2.4       &-2.1           &22.4         &5.4 \\
  \hline
   $\Delta \lambda_{PM}$  20-120$\,^{\circ}\mathrm{C}$  (nm)   &4.4   &-0.4       &-1.2           &29.1        &59.5 \\

  References              &\cite{Kato2002,  Konig2004, Fradkin1999,  Emanueli2003}    &\cite{Mikami2009, Mangin2011, Yutsis2004}                   &\cite{Emanueli2003, Fradkin-Kashi2000, Fenimore1995, Kato1994}                    &\cite{Kato2003, Cheng1994, Yutsis2004}                      &\cite{Mikami2011,  Cheng1993}  \\
  \hline  \hline
  \end{tabular}
  \end{adjustbox}
  \caption{Comparison of the chemical composition, $\lambda_{GVM1}$, $\Delta \lambda_{GVM1}=\lambda_{GVM1}(20\,^{\circ}\mathrm{C})-\lambda_{GVM1}(120\,^{\circ}\mathrm{C})$, $\lambda_{GVM1}$, $\Delta \lambda_{GVM2}=\lambda_{GVM2}(20\,^{\circ}\mathrm{C})-\lambda_{GVM2}(120\,^{\circ}\mathrm{C})$ and $\Delta \lambda_\textrm{PM}$ of PPKTP and four of its isomorphs.  The relevant sources in the literatures for the appropriate Sellmeier the thermal-optical equations are also listed in the table.}
  \label{table1}
\end{table*}
The GVM wavelength is an important parameter for the nonlinear crystals since it is the wavelength at which maximal spectral purity can be achieved.
There are three types of GVM conditions \cite{Edamatsu2011}:
\begin{equation}\label{eq:GVM1}
2V^{-1}_{g,p}(\lambda/2)=V^{-1}_{g,s}(\lambda)+V^{-1}_{g,i}(\lambda),
\end{equation}
\begin{equation}\label{eq:GVM2}
V^{-1}_{g,p}(\lambda/2)=V^{-1}_{g,s(i)}(\lambda),
\end{equation}
and
\begin{equation}\label{eq:GVM3}
V^{-1}_{g,s}(\lambda)=V^{-1}_{g,i}(\lambda),
\end{equation}
where $V^{-1}_{g,\mu} (\mu=p,s,i)$ is the inverse of the group velocity  $V_{g,\mu}$ for the pump $p$, the signal $s$,  and the idler $i$. $\lambda$ is the degenerate wavelength of the signal and idler. In our theoretical model, $p$, $s$ and $i$ propagate in a collinear configuration along the crystal's $x$ axis.
$p$ and $s$ polarize along the crystal's $y$ axis, while $i$ polarizes along the crystal's $z$ axis.
The group velocity can be calculated using the equation
\begin{equation}\label{eq4}
V_{g,\mu} \equiv \frac{d \omega} {dk}= \frac{c}{{n - \lambda dn}},
\end{equation}
where $c$ is the light velocity, $\omega$ is the angular frequency, $k$ is the wave vector and $n$ is the refractive index, which is a function of wavelength and temperature.
Each of the three GVM conditions has special applications in quantum state engineering.
For example, the first GVM wavelength ($\lambda_{GVM1}$) in Eq.\,(\ref{eq:GVM1}) can be used to prepare a round-shape joint spectral distribution \cite{Evans2010, Gerrits2011, Eckstein2011, Jin2013OE, Zhou2013, Bruno2014OE, Jin2016SR, Laudenbach2016, Konig2004, Li2015, Jin2017PRA}.
The second GVM wavelength ($\lambda_{GVM2}$) in Eq.\,(\ref{eq:GVM2}) can be used to prepare 90-degree-tilted oval-shape joint spectral distribution in KDP (KH$_2$PO$_4$) \cite{Mosley2008PRL, Jin2011}. The third GVM wavelength ($\lambda_{GVM3}$) in Eq.\,(\ref{eq:GVM3}) can be used to prepare narrow, long and 45-degree-tilted JSA in a PPSLT (periodically poled MgO-doped stoichiometric LiTaO$_3$) cystal \cite{Shimizu2009, Jin2016QST}.
In the case of \emph{M}TiO\emph{X}O$_4$ crystals, the third GVM condition can not be satisfied for type-II SPDC (although it can easily be satisfied for type-0 and type-I), therefore, we focus on the first two GVM conditions.

\begin{figure*}[th]
\centering\includegraphics[width=0.99\textwidth]{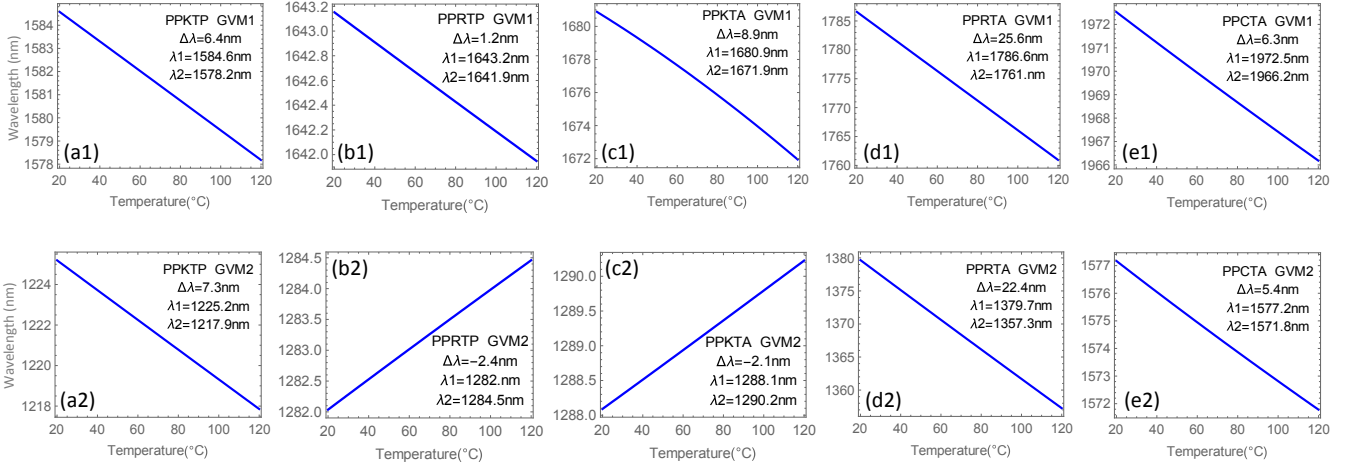}
\caption{$\lambda_{GVM1}$ (first row) and $\lambda_{GVM2}$ (second row) as a function of the temperature for different crystals. (a) PPKTP, (b) PPRTP, (c) PPKTA, (d) PRTA and (e) PPCTA. Inset: $\lambda_{1(2)}$ is  the wavelength at 20 (120)$\,^{\circ}\mathrm{C}$, and $\Delta\lambda=\lambda_{1}-\lambda_{2}$.
 } \label{gvm}
\end{figure*}

By using the Sellmeier equations and thermal-optical equations from the references as listed in Tab.\,\ref{table1}, we can calculate $\lambda_{GVM1}$ and $\lambda_{GVM2}$.
Fig.\,\ref{gvm}(a1-e1) shows the $\lambda_{GVM1}$ as a function of temperature from $20 \,^{\circ}\mathrm{C}$ to $120 \,^{\circ}\mathrm{C}$.
The $\lambda_{GVM1}$  is decreasing when the temperature is increasing, as shown in  Fig.\,\ref{gvm}.
Different crystals exhibit different amounts of decrease, as listed in the insets of Fig.\,\ref{gvm}. With $\lambda_{GVM1}$ shifting 25.6 nm, PPRTA has the biggest value of $\Delta \lambda$ while PPRTP shows the smallest wavelength shift of only 1.2 nm.
Fig.\,\ref{gvm}(a2-e2) depicts $\lambda_{GVM2}$ as a function of temperature. With increasing temperature, $\lambda_{GVM2}$ increases for PPRTP and PPKTA, while it decreases for the other crystals. PPRTA exhibits the biggest value of increase, i.e. 22.4 nm while PPKTA has the smallest value of 2.1 nm.

\begin{figure*}[th]
\centering\includegraphics[width=0.99\textwidth]{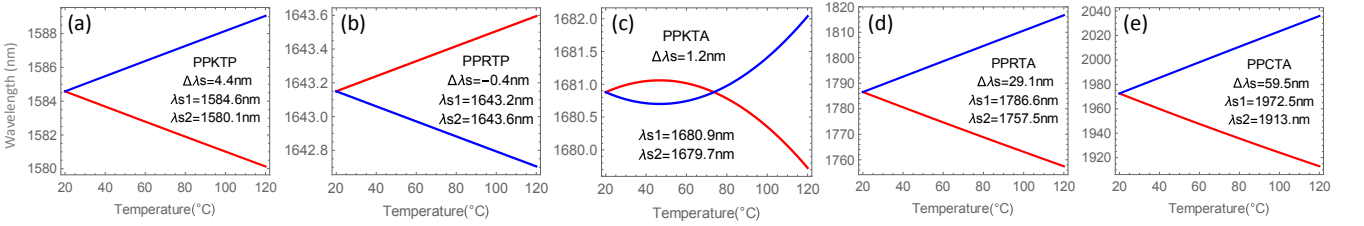}
\caption{The phase matched wavelength of signal (red line) and idler (blue line) as a function of temperature. (a) PPKTP, (b) PPRTP, (c) PPKTA, (d) PRTA and (e) PPCTA. Inset: $\lambda_{s1(2)}$ is  the wavelength of the signal at 20 (120)$\,^{\circ}\mathrm{C}$, and $\Delta\lambda_s=\lambda_{s1}-\lambda_{s2}$. In these plots, the poling period for each crystal is fixed to satisfy the degenerate type-II GVM1 condition at 20$\,^{\circ}\mathrm{C}$.
 } \label{phasemw}
\end{figure*}

\section{\emph{Thermal effect on the phase matching wavelength }}
Another important parameter for SPDC is the phase matching wavelength,
which can be obtained by solving the equation sets of
\begin{equation}\label{eq3-1}
\frac{1}{\lambda_p}-\frac{1}{\lambda_s}-\frac{1}{\lambda_i}=0,
\end{equation}
and
\begin{equation}\label{eq3-2}
k_p - k_s -k_i +\frac{2\pi}{\Lambda}=0,
\end{equation}
where  $\Lambda$ is the poling period. 
In fact, Eq.\,(\ref{eq3-1}) corresponds to the energy conservation law  and Eq.\,(\ref{eq3-2}) corresponds to the momentum conservation law during the SPDC process.

Note, the GVM wavelength $\lambda_{GVM}$ is different from the phase-matched wavelength $\lambda_{PM}$ , and these two parameters are independent from each other.
Generally speaking,  $\lambda_{PM}$ is the wavelength where SPDC occurs, while $\lambda_{GVM}$ is the wavelength where pure quantum state can be engineered.  $\lambda_{PM}$ is determined by the zero-order dispersion ($n(\lambda)$) of the crystal, while $\lambda_{GVM}$ is determined by the first order of dispersion ($\frac{dn(\lambda)}{d \lambda}$).
At a fixed temperature, the  $\lambda_{GVM}$ is fixed, but the $\lambda_{PM}$ can be shifted to any wavelength, as long as Eq.\,(\ref{eq3-1}) and Eq.\,(\ref{eq3-2}) are satisfied.

Figure\,\ref{phasemw}(a-e) show the phase matched wavelength as a function of the temperature from 20$\,^{\circ}\mathrm{C}$ to 120$\,^{\circ}\mathrm{C}$ for the \emph{M}TiO\emph{X}O$_4$.
Here, we set the pump wavelength at the half of $\lambda_{GVM1}$ at 20$\,^{\circ}\mathrm{C}$ and keep the poling period $\Lambda$ fixed.
In Fig.\,\ref{phasemw}, all the signal (red line) and idler (blue line) wavelength has a symmetric distribution around their GVM wavelength $\lambda_{GVM1}$. PPKTA shows a nonmonotonic change with the increase of temperature, while other crystals show a monotonic change. PPCTA is most sensitive to the temperature change. In contrast, PPRTP is very insensitive to temperature variance. With a temperature change from 20$\,^{\circ}\mathrm{C}$ to 120$\,^{\circ}\mathrm{C}$, the wavelength is changed by 59.5  nm for PPCTA, while only 0.4 nm for PPRTP.

\begin{figure*}[tbh]
\centering\includegraphics[width=0.99\textwidth]{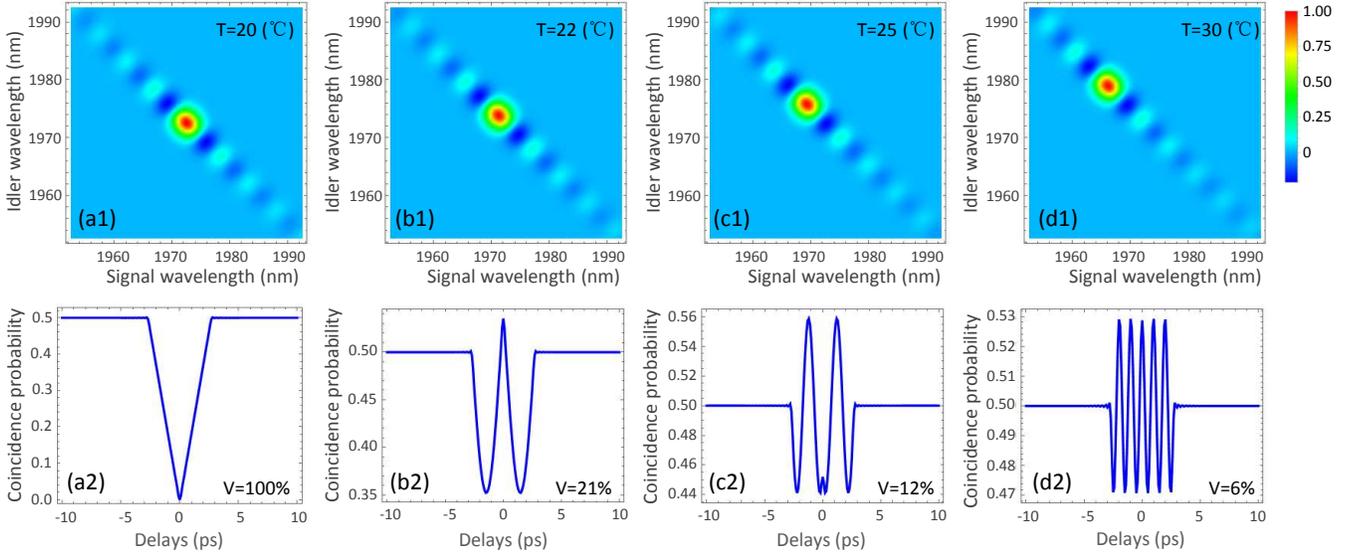}
\caption{ The JSAs (first row) and HOM interference patterns (second row) at different temperatures:  (a) 20$\,^{\circ}\mathrm{C}$, (b) 22$\,^{\circ}\mathrm{C}$, (c) 25$\,^{\circ}\mathrm{C}$ and (d) 30$\,^{\circ}\mathrm{C}$. In this simulation, the PPCTA length is 30-mm-long, while the pump laser bandwidth (full-width-at-half-maximum) is 0.87 nm.
 } \label{homi}
\end{figure*}

\section{\emph{Thermal effects on joint spectral distribution and Hong-Ou-Mandel interference  }}
With the Sellmeier and thermal-optical equation from the references as listed in Tab.\,\ref{table1}, we can also plot the joint spectral amplitude (JSA) as a function of temperature.
The JSA can be calculated using the following equation
\begin{equation}\label{eq1}
f(\omega _s ,\omega _i )=\phi (\omega _s ,\omega _i)\alpha (\omega _s +\omega _i )
\end{equation}
where $\phi (\omega _s ,\omega _i)$  and $ \alpha (\omega _s +\omega _i )$ are the phase matching amplitude and the pump envelope amplitude. 
Assuming the pump spectrum  has a Gaussian distribution with a bandwidth of $\sigma_p$, the pump envelope amplitude can be written as
\begin{equation}\label{eq1}
\alpha (\omega _s +\omega _i )= \exp [ - (\frac{{\omega _s  +\omega _i  - \omega _p }}{\sigma_p })^2 ].
\end{equation}
Under the collinear configuration in the SPDC process, the  phase matching amplitude can be written in the form of
\begin{equation}\label{eq1}
\phi (\omega _s ,\omega _i ) =  {\rm{sinc}}(\frac{{\Delta kL}}{2}),
\end{equation}
where $\Delta k = k_p  - k_s  - k_i + \frac{{2\pi }}{\Lambda }$, and
$L$ is the length  of the SPDC crystal.

The change in the JSA has an immediate effect on the Hong-Ou-Mandel interference \cite{Hong1987}.
The two-fold coincidence probability in a HOM interference can be calculated using the following equation \cite{Jin2015OE, Gerrits2015}.
\begin{equation}\label{eq:P2}
p(\tau ) =   \frac{1}{4} \int\limits_0^\infty  \int\limits_0^\infty  d\omega _s  d\omega _i \left| {[f(\omega _s ,\omega _i ) - f(\omega _i ,\omega _s )e^{ - i(\omega _s  - \omega _i )\tau } ]} \right|^2,
\end{equation}
where $\tau$ is the time delay in the interference.

Each of the  five crystals has different temperature-dependency in the HOM interference patterns. As an example, Fig.\,\ref{homi}(a1-e1)  shows the JSA of the signal and idler photons from PPCTA at the temperatures  20, 22, 25 and 30$\,^{\circ}\mathrm{C}$.
It can be observed that the JSA shifts to the top-left corner with the increase of the temperature.
As a result, the frequency entanglement between the signal and idler photons is shifted.
Such a frequency shift doesn't change the purity of the biphoton source, but severely affects the HOM interference patterns, as shown in Fig.\,\ref{homi}(a2-e2).
By increasing the temperature from 20 to 22, 25 and 30$\,^{\circ}\mathrm{C}$, the interference dips are  changed from 1 to 2, 3, 6. The visibility,  defined as $V=(P_{max}-P_{min})/(P_{max}+P_{min})$, is  decreased from 100\% to 21\%, 12\% and  6\%, where $P_{max(min)}$ is the maximal (minimal) value of $p(\tau)$.

\section{\emph{Discussion}}

It should be noticed that $\lambda_{GVM1}$ varies slightly depending on the respective  Sellmeier and thermal-optical equations measured by different groups, e.g., for PPKTP at room temperature, $\lambda_{GVM1}=1584.6$ nm and the poling period $\Lambda=45.0$ $\mu$m can be calculated from Ref.\, \cite{Kato2002}, while $\lambda_{GVM1}=1582.2$ nm and $\Lambda=46.1$ $\mu$m can be obtained by using the Sellmeier equation of $n_y$ from Ref.\,\cite{Konig2004} and $n_z$ from  Ref.\,\cite{Fradkin1999}.
For PPKTA, $\lambda_{GVM1}=1680.9$ and $\Lambda=50.2$ $\mu$m can be calculated using Refs.\,\cite{Emanueli2003, Fradkin-Kashi2000, Fenimore1995}, while   $\lambda_{GVM1}=1634.7$ and $\Lambda=57.3$ $\mu$m with Ref.\,\cite{Kato1994}.
For PPRTA, $\lambda_{GVM1}=1786.6$ and $\Lambda=73.3$ $\mu$m can be calculated using Ref.\,\cite{Kato2003}, while   $\lambda_{GVM1}=1784.5$ and $\Lambda=71.1$ $\mu$m with Ref.\,\cite{Cheng1994}.
For PPCTA, $\lambda_{GVM1}=1972.5$ and $\Lambda=248.4$ $\mu$m can be calculated using Ref.\,\cite{Mikami2011}, while   $\lambda_{GVM1}=1864.6$ and $\Lambda=381.9$ $\mu$m with Ref.\,\cite{Cheng1994}.
In the future, more experimental works are needed to obtain  more accurate Sellmeier equations and thermal-optical equations for these  crystals.

In Fig.\,\ref{homi}, the visibility of the HOM interference was degraded at a higher temperature.   In fact, the poling period $\Lambda$ of Eq.\,(\ref{eq3-2}) is a design parameter that is available for tuning the phase matching condition. Therefore, by varying $\Lambda$ appropriately for a specific temperature setting, the HOM visibility can be restored to unity by making sure that signal and idler wavelengths are the same.

The thermal-optical effect of quantum states generated from the isomorphs have many promising applications in the future.
Firstly, temperature can be used for precise control of the spectral distribution of the quantum state.
Sweeping the temperature can be used to scan the frequency.
This is very useful for  for spectroscopy \cite{Whittaker2017} or quantum interference \cite{Branczyk2011}.
Secondly, the crystals might be used for quantum sensing. PPCTA is very sensitive to temperature change, therefore it can be used as  temperature sensor  \cite{Esfahani2015}. 
In contrast, PPRTP is very insensitive to temperature changes, therefore, this crystal can be used at different temperatures, as an anti-temperature-variance crystal.

The KTP family has 118 known isomorphs \cite{Gazhulina2013, Sorokina2007, Stucky1989}, with a general formula of \emph{MM}$^{\prime}$O\emph{X}O$_4$, where \emph{M} = K, Rb, Na, Cs, Tl, NH$_4$; \emph{M}$^{\prime}$ = Ti, Sn, Sb, Zr, Ge, Al, Cr, Fe, V, Nb, Ta, Ga; \emph{X} = P, As, Si, Ge. The Sellmeier equations and thermal-optical equation for most of these crystals are not reported yet, therefore, exploring the thermal optical properties of these crystal is promising for further research. 

\section{Conclusion}
In conclusion, we have theoretically and numerically investigated the temperature-dependent properties of biphoton states generated from the isomorphs of the PPKTP crystal.
Specifically, the first and second type of GVM wavelength, the phase-matched wavelength, the JSA and HOM interference patterns as a function of temperature are explored in detail.
It is found that different crystals have different thermal effects. We summarize the thermal-dependent properties of the crystals in Tab.\,\ref{table1}.
The research results in this work should be very useful for the future applications, for example, in fine tuning of the spectrum in quantum interference or spectroscopy.

\section*{Acknowledgments}
This work is supported by the open research fund of Key Lab of Broadband Wireless Communication and Sensor Network Technology (Nanjing University of Posts and Telecommunications), Ministry of Education (Grant No. JZNY201709), by a fund from the Educational Department of Hubei Province, China (Grant No. D20161504), and by National Natural Science Foundations of China (Grant No. 61475075, 11104210 and 11704290).


%

\end{document}